# Analysis of Persian News Agencies on Instagram, A Words Co-occurrence Graph-based Approach

Mohammad Heydari[a], Babak Teimourpour*[b]

Faculty of Industrial and Systems Engineering, Tarbiat Modares University, Tehran, Iran; m_heydair@modares.ac.ir[a], b.teimourpour@modares.ac.ir[b]

## *ABSTRACT*

**The rise of the Internet and the exponential increase in data have made manual data summarization and analysis a challenging task. Instagram social network is a prominent social network widely utilized in Iran for information sharing and communication across various age groups. The inherent structure of Instagram, characterized by its text-rich content and graph-like data representation, enables the utilization of text and graph processing techniques for data analysis purposes. The degree distributions of these networks exhibit scale-free characteristics, indicating non-random growth patterns. Recently, word co-occurrence has gained attention from researchers across multiple disciplines due to its simplicity and practicality. Keyword extraction is a crucial task in natural language processing. In this study, we demonstrated that high-precision extraction of keywords from Instagram posts in the Persian language can be achieved using unsupervised word co-occurrence methods without resorting to conventional techniques such as clustering or pre-trained models. After graph visualization and community detection, it was observed that the top topics covered by news agencies are represented by these graphs. This approach is generalizable to new and diverse datasets and can provide acceptable outputs for new data. To the author's knowledge, this method has not been employed in the Persian language before on Instagram social network. The new crawled data has been publicly released on GitHub for exploration by other researchers. By employing this method, it is possible to use other graph-based algorithms, such as community detections. The results help us to identify the key role of different news agencies in information diffusion among the public, identify hidden communities, and discover latent patterns among a massive amount of data.**

## 1. Introduction

Instagram is a prominent social network renowned for its role in facilitating information sharing, communication and holds significant importance as one of the most widely used platforms among diverse age groups in Iran [1]. The study of word co-occurrence plays a crucial role in various natural language processing (NLP) tasks, such as word embedding, sentiment analysis, and information retrieval.

The analysis of word relationships and contextual associations can provide valuable insights into the semantic and syntactic properties of language. Co-occurrence networks are commonly employed to visualize potential connections between individuals, organizations, and other entities.

The production and visualization of co-occurrence networks have become possible with the emergence of electronic text storage and exploration [2].

A co-occurrence network is a collection of connections between terms or words, based on their presence together in a specific section of a text. These networks are constructed by assessing the relationship between each pair of words using a co-occurrence index. For instance, if two words, A and B, both appear in a particular article, they are considered co-occurrences. Similarly, if another article contains B and C together, a connection is established between B and C. By connecting A to B and B to C, a co-occurrence network is formed using these words. The method of defining co-occurrence within a textual segment can vary. One of the fundamental applications of co-occurrence networks is extracting keywords and detecting trends. Keyword extraction is a critical task in the field of natural language processing [3].

Keywords serve as concise summaries of text content. Previously, this task was performed and analyzed manually by humans, but today it is accomplished through automated processes [4].

In this study, we first review recent efforts and progress in the field of keyword extraction and word co-occurrence network analysis. Next, we describe how the data was extracted and provide general explanations regarding the quantity and quality of the data utilized. The subsequent section briefly outlines the preprocessing steps applied to the textual data. Following that, we delve into the analysis of the co-occurrence network of words in these texts. Centralities, communities, and degree distribution are among the analytical approaches employed in this study. Subsequently, we present and elucidate the visualization of this network. Finally, we draw conclusions and offer suggestions for future research.

## 2. Related Works

Mirasadr proposed WordGraph2Vec [5], a graph-based word embedding algorithm that uses a large corpus to create a word co-occurrence graph. It then samples word sequences from this graph by random traversal and trains word embeddings on these samples. The algorithm benefits from the stable vocabulary and expressions in English, resulting in minimal changes to the graph's size and density with increasing training corpus. This stability enables Word-Graph2vec to have consistent runtime on large datasets and increasingly outperform traditional Skip-Gram models as the corpus size grows. The authors experimental results demonstrate that Word-Graph2vec achieves significantly higher efficiency (four to five times) while maintaining low error rates through random walk sampling.

In another study, Mirasadr et al. [5] discussed Comprehensive and detailed about Graph of Words Model for Natural Language Processing. Muskan studied on an exploration of various aspects concerning techniques, problems and difficulties for extracting graphical keywords [6]. In linguistics and natural language processing, words co-occurrence or cooccurrence topic is currently of great importance, and many academics are interested in applying this method in their work.

Fudolig et al. [7] examines the relationship between context and happiness scores in political tweets by analyzing word co-occurrence networks. Nodes in the network represent words, and the edge weights indicate the frequency of co-occurrence in the tweet corpus. When community detection is applied to the network, meaningful word groups with distinct themes emerge, and the happiness scores of the words within each group correspond to their respective theme.

Futrel [8] focused on utilizing pretrained static word embeddings to improve the estimation of bilexical co-occurrence probabilities. These probabilities, which represent the likelihood of one word given another in a specific relationship, are crucial in psycholinguistics, corpus linguistics, and cognitive modeling of language. The proposed approach involves a log-bilinear model that takes pretrained vector representations of both words as input, allowing for generalization based on the distributional information captured by the vectors.

Gabin et al. [9] introduced two innovative models for keyword suggestion, trained specifically on scientific literature. Their methods adapt the architectures of Word2Vec and FastText to generate keyword embeddings, utilizing the co-occurrence of keywords in documents. To evaluate the performance of their models, they developed a ranking-based evaluation methodology that encompasses both known-item and ad-hoc search scenarios.

Vega-Oliveros et al. [10] introduced a method called the multi-centrality index (MCI) approach. Its purpose is to determine the best combination of word rankings based on different centrality measures. They investigate nine centrality measures (Betweenness, Clustering Coefficient, Closeness, Degree, Eccentricity, Eigenvector, K-Core, PageRank, Structural Holes) for the identification of keywords in document representations using co-occurrence word-graphs.

Biswas et al. [11] introduces a new unsupervised graph-based method for keyword extraction called KECNW. It assesses keyword importance by considering multiple influencing parameters. The KECNW utilizes Node Edge rank centrality with node weight determined by various factors. The effectiveness of the model is evaluated using five datasets: Uri Attack, American Election, Harry Potter, IPL, and Donald Trump.

Li et al. [12] proposed a graph-based word embedding algorithm, called Word-Graph2vec, which converts the large corpus into a word co-occurrence graph, then takes the word sequence samples from this graph by randomly traveling and trains the word embedding on this sampling corpus in the end.

Hettiarachchi et al. [13] introduced a new approach called WhatsUp, which aims to identify precise temporal and detailed textual event information. This method utilizes linguistics through self-learned word embeddings, as well as hierarchical relationships and frequency-based measures for statistical analysis.

Nazar et al. [14] proposed an automated approach for specialized term extraction. They used a statistical measure based on term co-occurrence to assess semantic relevance within a specific domain. By prioritizing term candidates that frequently co-occur with a particular set of units, they assign higher ranks to them. No external resources are required, but performance benefits from a pre-existing term list.

Valentini et al. [15] studied gender bias measurement using word embeddings. They discovered that Skip-gram with negative sampling and GloVe algorithms tend to identify male bias in high-frequency words, whereas GloVe shows a tendency to indicate female bias in low-frequency words.

To enhance the effectiveness of the TextRank [16] algorithm, Xiong et al. [17] introduce a novel approach called Semantic Clustering TextRank (SCTR). This algorithm is designed for extracting news keywords and leverages the power of TextRank while incorporating semantic clustering. Initially, the word vectors produced by the Bidirectional Encoder Representation from Transformers (BERT) model [18] are employed to execute k-means clustering, thereby establishing semantic clusters. Subsequently, the outcomes of this clustering process are utilized to construct a TextRank weight transfer probability matrix. Lastly, an iterative procedure is conducted to calculate word graphs and extract the keywords from the text.

Ling et al. [19] proposed a lightweight algorithm for keyphrase extraction that doesn't require external resources. Their method ranks words and generates keyphrases by analyzing sentence semantics. The algorithm initializes and updates word values iteratively, incorporating sentence information. This approach improves accuracy and reduces the number of iterations. They also propose an unsupervised keyword extraction technique for efficient retrieval of article topics.

Leilei et al. [20] proposed SemGloVe, which distills semantic co-occurrences from BERT into static GloVe word embeddings. Exclusively, they suggest two models to extract co-occurrence statistics based on either the masked language model or the multi-head attention weights of BERT. Their technique can extract word pairs limited by the local window hypothesis and can explain the co-occurrence weights by directly considering the semantic distance between word pairs. Experiments on several word similarity datasets and external tasks show that SemGloVe can surpass GloVe. Anjali et al. [21] proposed a graph-based model approach for extracting

keywords from a research paper. The proposing method is a combination of Rapid Automatic Keyword Extraction (RAKE) algorithm and Keyword Extraction using Collective Node Weight (KECNW), in which candidate keywords are extracted using RAKE algorithm and selection of keywords using KECNW model.

Liujun et al. [22] proposed an unsupervised keyword extraction method based on text semantic graphs. They improved the construction of the text graph and word weight calculation. By considering the semantic dependencies of words, they built a text semantic graph with four types of edges. This eliminates the need for parameter setting in the traditional graph generation method. Word weights were calculated based on keyword position, term frequency-inverse document frequency, concept level, and connection strength. The importance of words was sorted, and high-score nodes were selected to form the keyword set for the abstract text.

Brock et al. [23] introduced Textstar, a graph-based system for summarization and keyphrase extraction. Textstar constructs a document graph using basic lemmatization and POS tagging, connecting lemmas and sentences. Through an iterative centrality algorithm, low-ranked nodes are progressively removed to generate summaries. The remaining sentences form the summary, while the remaining lemmas are combined into key phrases based on their context. The summary of related works is shown in Table 1.

## 3. Data

### 3.1 Data Crawling

Initially, the research study involved web crawling of several Instagram pages, namely BBC Farsi, Fars News Agency, and KhabaFouri based on Python programming language and Selenium library. It is important to note that the original data of these pages consisted of 14,000 posts from BBC Farsi, 25,000 posts from Fars News Agency, and 17,000 posts from KhabarFouri. However, for the purpose of this study, a subset of 2000 posts were selected on average from each page. The data crawling time is from October 2019 to February 2020. The collection of data was not based on any special subjective assumption and was only collected in the mentioned time.

The Figure 1, Figure 2 and Figure 3 contains data related to three different sources (Fars, BBC, and Khabarfouri) and their respective posts on Instagram platform. These statistics provide insights into the posting behavior and content length of each source. The "Average Number of Words per Post" is particularly useful for understanding the length of the posts published by each source, with BBC's articles being the longest on average, followed by Fars and Khabarfouri.

### 3.1 Data Preprocessing

After the data was crawled and aggregated, the texts were processed in terms of their composition or writing style. In this class, the following tasks or activities are performed: The list provided outlines various text processing tasks commonly performed in natural language processing and text analytics.

1) Removing emojis and punctuations: eliminating emojis and punctuations from text data to focus on the textual content.

2) Removing links, addresses, IDs, and hashtags: as they may not provide meaningful information for analysis.

3) Removing Stop words: that do not carry significant semantic meaning, such as prepositions, and conjunctions.

4) Converting Persian numbers to English: Transforming Persian numeral representations into English equivalents for consistency and ease of analysis. That's Because all the numbers become the same and become normal.

5) Spacing between numbers and letters: Inserting spaces between numbers and letters for better readability and consistency in the text.

6) Removing all characters except Persian language characters and numbers: Retaining only Persian language characters (letters) and numbers, while discarding any non-Persian or special characters.

7) Removing all text components except attributes and names

8) normalize texts using Hazm library.

Table 1. Classification of Related Works

| Author | Major Ideas | Strengths |
|---|---|---|
| Mirasadr | WordGraph2Vec for word embeddings | High efficiency, minimal changes in graph size |
| Muskan | Exploration of keyword extraction techniques | Addresses a current topic of importance |
| Fudolig | Analyzing word co-occurrence in political tweets for assessing happiness scores | Emergence of meaningful word groups, correspondence to themes |
| Futrel | Using pretrained word embeddings to estimate bilexical co-occurrence probabilities | Improves estimation of co-occurrence probabilities |
| Gabin | Innovative keyword suggestion models trained specifically on scientific literature | Adapts Word2Vec and FastText architectures for keyword embeddings, novel evaluation |
| Vega-Oliveros | Multi-centrality index approach for keyword identification | Investigates various centrality measures for keyword identification |
| Biswas | Unsupervised keyword extraction (KECNW) method with node-edge rank centrality | Considers multiple influencing parameters for keyword importance |
| Li | Word-Graph2Vec for word embeddings | Efficient training, word graph generation |
| Hettiarachchi | WhatsUp method for precise event information identification | Utilizes linguistics, hierarchical relationships for statistical analysis |
| Nazar | Automated specialized term extraction method based on term co-occurrence | Utilizes term co-occurrence for semantic relevance assessment |
| Valentini | Studying gender bias in word embeddings | Reveals gender bias tendencies in word embeddings |
| Xiong | Semantic Clustering TextRank (SCTR) for keyword extraction | Incorporates semantic clustering, TextRank for keyword extraction |
| Ling | Lightweight keyphrase extraction algorithm without external resources | Improved accuracy, efficient topic retrieval and unsupervised keyword extraction |
| Leilei | SemGloVe for distilling semantic co-occurrences from BERT into GloVe word embeddings | Surpasses GloVe in word similarity tasks and external tasks |
| Anjali | Combining RAKE and KECNW for keyword extraction from research papers | Combines two extraction models for keywords |
| Liujun | Unsupervised keyword extraction based on text semantic graphs | Improved text graph construction, word weight calculation, and semantic distance |
| Brock | Textstar for summarization and keyphrase extraction | Summarization and keyphrase extraction based on document graph |

The Hazm library's normalizer performs the following functions: a) Correcting spaces and half spaces and b) Correcting and unifying polymorphous words.

Following that, the text documents, represented by the captions under the posts, are treated as individual units. As per common practice in many articles, it is essential to construct a co-occurrence matrix for the documents in the corpus. This involves treating each post on the Instagram pages as a separate document and updating the overall matrix by considering the word co-occurrence within each document.

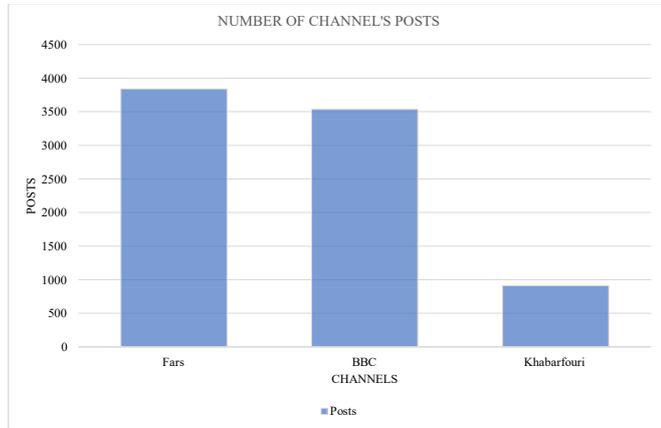

Figure. 1. Number of Channel's Posts

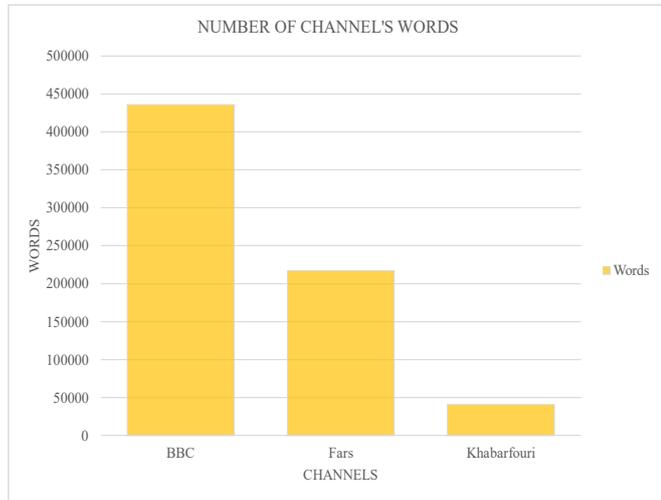

Figure. 2. Number of Channel's Words

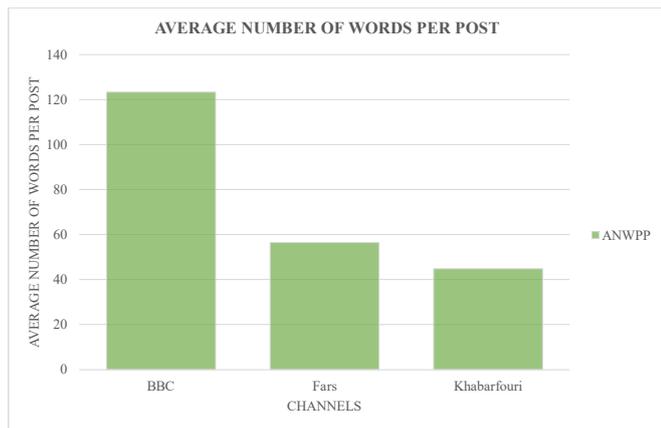

Figure. 3. Average number of words per post

For this purpose, a sentence splitter is employed to segment the texts into individual sentences. Then, using a Part-of-Speech (POS) tagger, all sentence components except adjectives and nouns are removed. This approach, commonly found in various keyword extraction articles, aims to refine the input for more precise and effective keyword extraction. After applying the sentence splitting and POS tagging process, the resulting documents are once again saved as document files in a specified format or file structure.

This allows for further analysis or processing of the documents in their updated form. As mentioned, it is important to consider the text corpus in the form of complete documents rather than viewing individual sentences separately when creating a co-occurrence matrix.

By treating each document as a cohesive unit, the co-occurrence relationships between both words and sentences can be captured and analyzed effectively in the resulting matrix. After normalizing the text, the co-occurrence matrix of the words was constructed using the following process. Firstly, a single dictionary containing all unique words from the normalized text was created. Subsequently, a raw word co-occurrence matrix was generated. This matrix was initially empty and would be filled by iterating through the data. In this context, the documents referred to here are equivalent to the posts. During the traversal of the data, the co-occurrence matrix was populated based on the presence of word pairs within the same document. The weight associated with each word increased when it co-occurred with another word in the same document. By considering such co-occurrence relationships, the matrix effectively captured the associations between words in the context of the documents. The Table 2 is a sample part of the whole dataset.

## 4. Network Science Techniques

In third section of our research, we employed graph analytic techniques to discover latent patterns of network and detect hidden communities in the network [24]. Before proceeding further, we analyze and identify significant centralities within the network, which will provide us with valuable insights into the data.

### 4.1. Closeness Centrality

The closeness centrality referred to in connected graphs is often defined as the inverse of the shortest distance between a node and other nodes in the graph. In the context of this study, it can be inferred that words with higher centrality have been positioned closer to other words more frequently or have been utilized in the text of news articles more frequently. In other words, words with higher centrality are more central or influential within the co-occurrence network of words. They tend to have stronger connections and interactions with other words, indicating their significance in the context of the news texts. Graphs closeness and Betweenness centrality values are shown in Tables 3-8 respectively.

Table 2. Part of the word co-occurrence matrix

| قیمت | نمایندگی | بنزین | فوتبال | رادیو | سهمیه | * |
|---|---|---|---|---|---|---|
| ۳ | ۱ | ۸ | ۲ | . | . | سهمیه |
| . | . | . | . | . | . | رادیو |
| . | . | . | . | . | ۲ | فوتبال |
| ۱۱ | . | . | . | . | ۸ | بنزین |
| . | . | . | . | . | ۱ | نمایندگی |
| . | . | . | ۱۱ | . | ۳ | قیمت |

## 4.2. Betweenness Centrality

The betweenness centrality measure that considers all the shortest paths in a graph is known as betweenness centrality. It calculates the importance of a node in the network based on the number of shortest paths that pass through that node.

## 4.3 Community Detection

One of the intriguing findings in this research was the identification of word communities within the news texts. By analyzing the words and their associations within the texts, we can group them into communities where words within each community exhibited a strong interconnectedness. These communities represent groups of words that are closely related. Considering the significant number of communities formed, below are some examples of important communities observed within each page:

### *BBC Farsi:*
- Community 1

Words related to "gasoline prices," "economy," and "inflation." The cluster shows that the price of gasoline has a significant effect on the inflation rate in Iran's economy. So, a one percent increase gasoline price increases the inflation rate by 1.112 percent. Also, the price of gasoline has a significant effect on the inflation rate in Iran.

- Community 2

Words associated with "protests," "demonstrations," and "social unrest." The concept of the first cluster reveals that the rise in gasoline prices, which contributes to inflation and economic stagnation, has consequences that cannot be ignored. It is likely to result in public demonstrations as a manifestation of dissatisfaction.

### *Fars News Agency:*
- Community 3

Words pertaining to "Iran-U.S. tensions," "conflicts," and "diplomatic relations." The concepts in this cluster indicate the existing tensions between Iran and America, political conflicts, and ambiguous diplomatic relations.

- Community 4

Words linked to "regional politics," "Middle East," and "international affairs." The concepts within this cluster are influenced by the first cluster in the political domain, which pertains to Iran's regional policies, the volatile situation in the Middle East, and the longstanding interest of the West in acquiring oil resources, and strengthening ties with Persian Gulf nations such as Saudi Arabia, the UAE, and Qatar.

Table 3. Closeness centrality in BBC Farsi Instagram channel

| Nodes | Nodes | Closeness Centrality |
|---|---|---|
| Iran | ایران | 0.2122 |
| Iranian Majles | مجلس | 0.1818 |
| Iraq | عراق | 0.1774 |
| Republic | جمهوری | 0.1769 |
| USA | آمریکا | 0.1750 |
| Islamic | اسلامی | 0.1714 |
| Government | دولت | 0.1668 |

Table 4. Closeness centrality in Fars News Instagram channel

| Nodes | Nodes | Closeness Centrality |
|---|---|---|
| Iran | ایران | 0.1651 |
| USA | آمریکا | 0.1603 |
| City | شهر | 0.1504 |
| Foreign | خارجه | 0.1496 |
| Attack | حمله | 0.1469 |
| Representative | نماینده | 0.1443 |
| Trump | ترامپ | 0.1429 |

Table 5. Closeness centrality in Khabarfouri Instagram channel

| Nodes | Nodes | Closeness Centrality |
|---|---|---|
| Price | قیمت | 0.0631 |
| Toman | تومان | 0.0539 |
| Dollar | دلار | 0.5208 |
| Gasoline | بنزین | 0.4639 |
| Increasement | افزایش | 0.4639 |
| Currency | ارز | 0.4431 |
| Coin | سکه | 0.4366 |

Table 6. Betweenness centrality in BBC Farsi Instagram channel

| Nodes | Nodes In Persian | Betweenness Centrality |
|---|---|---|
| Iran | ایران | 0.7472 |
| Chief Commander | فرمانده | 0.7287 |
| Iranian Majles | مجلس | 0.5679 |
| Islamic | اسلامی | 0.4989 |
| Revolution | انقلاب | 0.4465 |
| Minister | وزیر | 0.0406 |
| News Agency | آژانس | 0.3978 |

Table 7. Betweenness centrality in Fars News Instagram channel

| Nodes | Nodes In Persian | Betweenness Centrality |
|---|---|---|
| Iran | ایران | 0.1089 |
| City | شهر | 0.0778 |
| USA | مجلس | 0.0748 |
| Tehran | اسلامی | 0.0644 |
| Government | دولت | 0.0573 |
| Minister | وزیر | 0.0417 |
| Council | شورا | 0.0408 |

Table 8. Betweenness centrality in KhabarFouri Instagram channel

| Nodes | Nodes In Persian | Betweenness Centrality |
|---|---|---|
| Tehran | تهران | 0.0133 |
| Sabotage | خرابکاری | 0.0124 |
| Price | قیمت | 0.0106 |
| Cost | هزینه | 0.0089 |
| Representative | نماینده | 0.0080 |
| Country | کشور | 0.0072 |
| Iran | ایران | 0.0071 |

*KhabarFouri:*
- Community 5

Words connected to "Iraq protests," "political instability," and "government." Latent concepts in this cluster refers to Iraqi protests, starting on October 1, 2019, a series of demonstrations, marches, sit-ins, and acts of civil disobedience occurred in Iraq, lasting until 2021. Protests were initially organized by civil activists through Instagram, primarily in the central and southern provinces of the country. The main grievances expressed by the protesters included corruption, high unemployment rates, political sectarianism, inadequate public services, and foreign interference. The movement quickly spread to other provinces in Iraq, with coordination facilitated through social media channels. By late October, the intensity of the demonstrations reached its peak, and protesters not only demanded a complete restructuring of the Iraqi government but also sought to diminish Iranian influence, including the presence of Iranian-aligned Shia militias.

- Community 6

Words associated with "domestic policies," "social issues," and "public opinion." After conducting a comprehensive analysis of the data extracted from final cluster, the following socio-economic and political challenges emerge as the most significant in Iran: The government's administration of the economy, particularly during the 1970s, The country's economic dependence on the dollar, Lack of dynamism in foreign trade, Stringent sanctions imposed by the United States, Inadequate economic patterns within society, including excessive extravagance, luxury enjoyed by a privileged few, and a lack of preference for domestic products.

These are just a few examples of the important communities identified within each page. Each community represents a group of words that frequently co-occur and share strong associations within the news texts, indicating their thematic or semantic relevance to specific topics or events. One of the interesting results is the community search in words, after examining the words in the news texts and socializing them, the results of each community and in fact the words that were placed next to each other, were related to each other to a great extent. Due to the large number of communities (clusters) created, some of the important communities of each page are listed below. To assess whether the networks are random or follow a scale-free distribution, the degree distribution of each network is examined. The degree of a node in a network refers to the number of connections it has with other nodes. In the following of our study, Figure 4, Figure 5 and Figure 6 demonstrates degree distribution of BBC Persian, FarsNews and KhabarFouri graphs.

The observation that the degree distribution graphs resemble a scale-free graph indicates that these networks do not exhibit random characteristics. Instead, they belong to the category of growing graphs. The term "growing graphs" suggests that these networks have evolved over time, with certain nodes acquiring more connections compared to others.

The Network Diameter is a measure of the maximum distance, in terms of the number of edges or steps, between the two most distant nodes within a network. It reflects the longest shortest path in the network and is an essential metric for understanding the network's overall size and connectivity. Based on Figure 7, Here are what values indicate for three graphs: A network diameter of 15 suggests that, within the BBC network, the longest shortest path between any two nodes is 15 edges. In other words, the most distant nodes in the BBC network can be reached by traversing a path of no more than 15 connections or steps. The KhabarFouri network has a larger diameter of 25. This indicates that it takes up to 25 edges or steps to travel from the farthest node to another in this network. The KhabarFouri network is larger or less tightly connected than the BBC network in terms of network diameter. The FarsNews network has a network diameter of 18. This means that, within the FarsNews network, the longest shortest path between any two nodes is 18 edges. It is somewhat more connected or smaller in diameter compared to the KhabarFouri network but larger than the BBC network. In summary, the network diameter values provide insights into the overall size and connectivity of these networks. A larger diameter suggests a network with nodes that are, on average, more distant from each other, while a smaller diameter implies a more tightly connected network with nodes that are closer to each other in terms of network traversal.

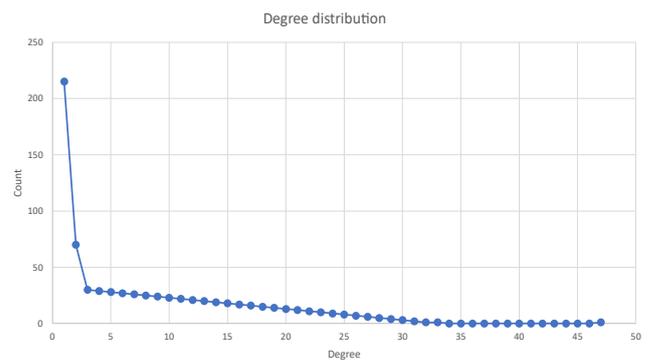

Figure. 4. Degree distribution on the BBC Farsi Instagram Channel

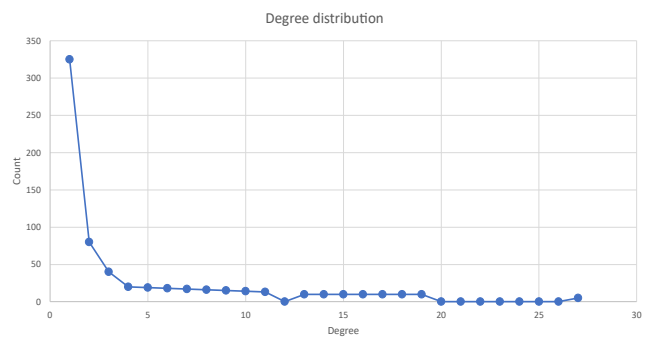

Figure. 5. Degree distribution on Fars news Instagram Channel

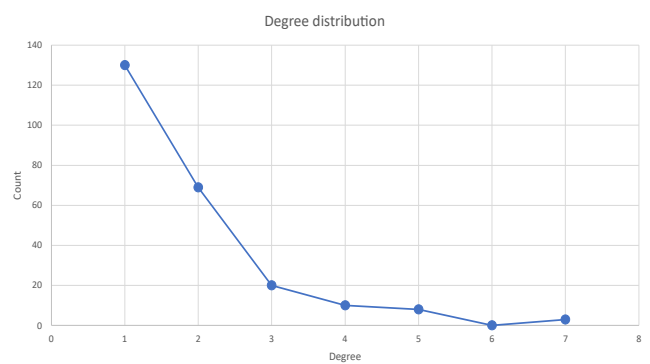

Figure. 6. Degree distribution on KhabarFouri Instagram Channel

Based on the Figure 8, The average degree for the BBC graph is 1.023. This indicates that, on average, each node (representing a specific word) in the BBC graph is connected to approximately 1.023 other nodes. The average degree for the FarsNews graph is 0.962. This means that, on average, each node in the FarsNews graph has connections to about 0.962 other nodes and finally In the case of the KhabarFouri graph, the average degree is 0.807. This suggests that, on average, each node in the KhabarFouri graph is linked to approximately 0.807 other nodes. These average degrees provide insight into the connectivity and relationships between nodes in each graph, with higher average degrees indicating a higher degree of connectivity among the entities or items represented in the graph.

The modularity score is a measure of the quality of community structure within a network or graph. A higher modularity score indicates a better division of nodes into distinct communities or clusters. For the community detection we utilized Louvain algorithm. Based on Figure 9, the modularity scores for three graphs: The modularity score for the BBC graph is 0.784. This score suggests that the nodes in the BBC graph are moderately well-organized into communities, but there might be room for improvement in terms of identifying distinct groups of related nodes. The KhabarFouri graph has a modularity score of 0.918. This is a high modularity score, indicating a strong and well-defined community structure within the graph. Nodes in this graph are highly organized into separate clusters and finally the FarsNews graph has a modularity score of 0.847. This score is also relatively high, suggesting a good division of nodes into distinct communities, but it may not be as strong as the community structure in the KhabarFouri graph. In summary, both the KhabarFouri and FarsNews graphs exhibit a strong community structure, with the KhabarFouri graph having the highest modularity score, indicating a well-defined organization of nodes into communities. The BBC graph has a slightly lower modularity score, indicating a moderate level of community structure.

The average path length is a metric that measures the average number of steps or hops it takes to travel from one node to another in a graph. It can provide insights into the overall connectedness and efficiency of information or influence flow within a network. Based on Figure 10, average path lengths for three graphs: The average path length for the BBC graph is 4.222. This relatively low average path length suggests that, on average, it takes about 4.222 steps to navigate from one node to another within the BBC graph. This indicates a relatively efficient information or influence flow within the BBC network. The average path length for the FarsNews graph is 6.352. A higher average path length, such as this, suggests that it takes more steps, approximately 6.352 on average, to move from one node to another within the FarsNews graph. This indicates a less efficient or more spread-out network structure. The KhabarFouri graph has an average path length of 5.384. This is higher than the average path length of the BBC graph but lower than that of the FarsNews graph. It suggests that, on average, it takes about 5.384 steps to traverse from one node to another within the KhabarFouri graph. The KhabarFouri network falls between the BBC and FarsNews graphs in terms of network efficiency. In summary, the average path length can be used to assess how efficiently information or influence can spread through the network. The BBC graph has the lowest average path length, indicating a relatively efficient network structure, while the FarsNews graph has the highest average path length, suggesting a less efficient or more widely dispersed network.

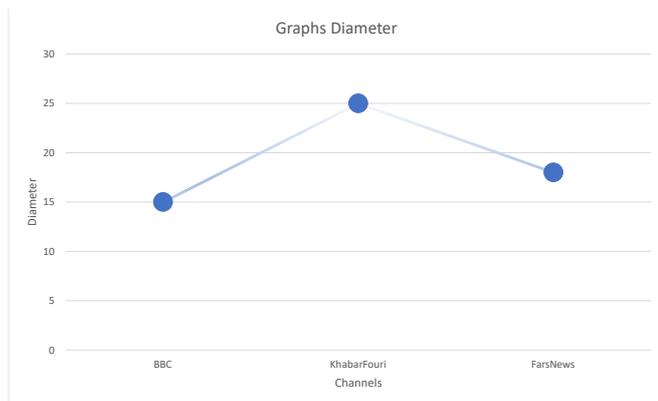

Figure. 7.   Various Networks Diameter

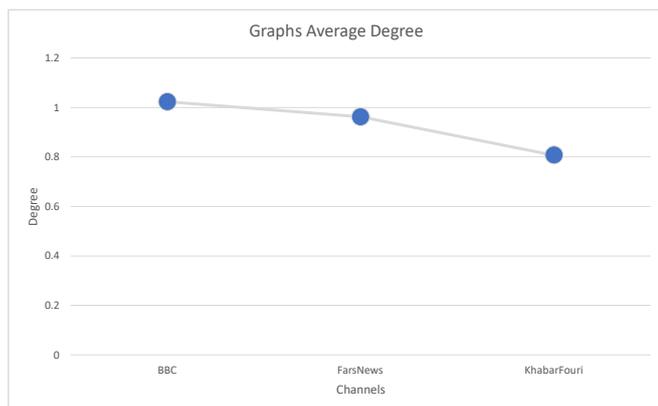

Figure. 8.   BBC Persian, Fars News and KhabarFouri Average Degree

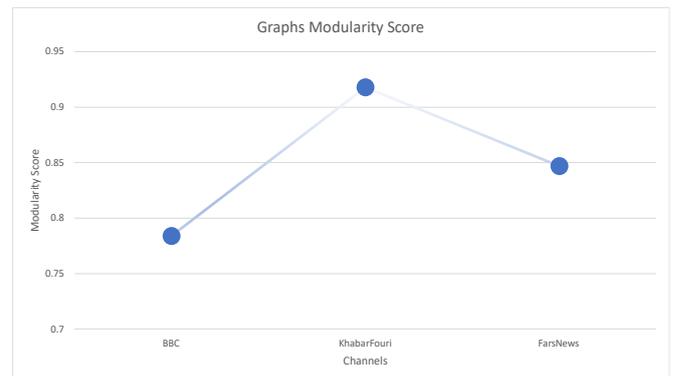

Figure. 9.   BBC Persian, Fars News and KhabarFouri Modularity Score

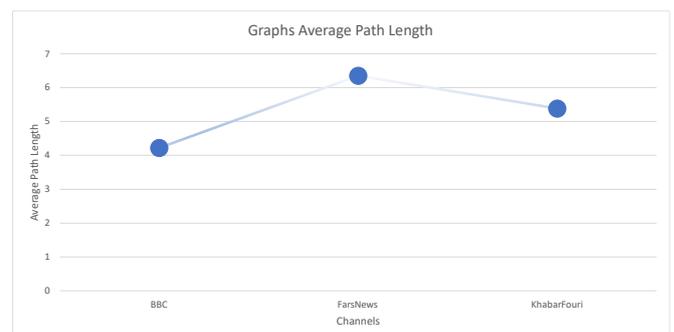

Figure. 10. BBC Persian, Fars News and KhabarFouri Average Path Length

**4.4 Network Visualization**

The visualization part of the graphs derived from the relationship matrix is highly significant that enables a deeper understanding of the obtained network and their underlying structures. Therefore, it becomes evident that each group of words forms distinct and meaningful clusters. These clusters represent coherent topics or themes that will be further discussed and analyzed.

The analysis of the graph reveals that Iran is the central focus of the discussed words, indicating its prominence in the interconnected topics. The graph is divided into different communities, visually represented by distinct colors, which signifies that related words tend to group together. This enables us to identify the prevailing topics and issues that have gained attention in recent months. One noticeable theme highlighted in the graph is the martyr Qassem Soleimani burial in Kerman, shedding light on its significance and the discussions surrounding it. The Iranian Revolutionary Guards and the Qods Force, both prominent entities in Iran, are also emphasized, indicating their relevance within the context of the graph. Other significant topics that emerge from the graph include references to the Ukrainian plane crash, the family members of Pooya Bakhtiari, the President, gasoline prices, casualties in Tehran, and internet-related matters. Even though certain hashtags may appear less prominent on the graph and are often overlooked because of their peripheral positioning, it is crucial not to underestimate their significance.

In specific timelines, these very nodes have played a crucial part in disseminating information across the network. Examples include their involvement in spreading news about the plane crash and the nuclear agreement. The clustering of these words within their respective communities suggests a strong interconnectedness among them, revealing their relevance and the relationships they share. The modularity value of 0.552 serves as a metric to assess the division of the network into meaningful communities. In this case, the relatively high modularity value indicates a well-defined and coherent division of the graph into distinct clusters. This reinforces the notion that the identified communities are significant and reflect the interconnected nature of the discussed issues within the scope of Iran. Overall, the graph provides valuable insights into the prevailing topics and trends of recent months, highlighting the interconnectedness of various issues in Iran.

The expressions mentioned, such as the reduction of JCPOA obligations, payment of cash subsidy, European Union foreign policy official, Haj Qasem, rationing, and increase in gasoline prices, align with the focus on Iran and its related issues discussed in the graph. These expressions represent specific topics or events that have gained attention and are interconnected within the network. Regarding the comparison between the Fars News Agency and the BBC in terms of the number and type of gatherings, it can be inferred that the Fars News Agency covers a broader range of topics compared to the BBC. This conclusion is based on the observation that the Fars News Agency exhibits a higher diversity and variety of news topics, as indicated by the communities and clusters identified within the graph.

The text points out that although specific expressions like the price of gasoline, the secretary of the Airline Companies Association, equity dividends, British Prime Minister Boris Johnson, and Shabnam Nematzadeh are not directly shown in the graph's description and communities, it emphasizes that the graph provides a broader view of the popular topics and the relationships between words associated with Iran. Final visualization of three graphs are shown in Figures 11-13 respectively.

**5. Conclusion**

This research has unveiled the power of word co-occurrence networks in extracting keywords and revealing latent patterns within Instagram text data from Persian news agencies. Through data crawling and preprocessing, we prepared the textual corpus for analysis. Network science techniques, including centrality metrics and community detection, uncovered significant word clusters and their interconnectedness, leading to the identification of key topics and trends. The strength of the methods based on the large language models (LLM) is the massive data set of the pre-training model and the size of the model that it is comprehensive in every aspect of the model, but in this study, we have achieved acceptable results with a relatively small dataset that should not be overlooked. This research demonstrates that precise keyword extraction from Persian-language Instagram posts can be effectively accomplished using unsupervised word co-occurrence techniques, eliminating the need for traditional methods like clustering or pre-trained models. The study involved graph visualization and community detection, revealing that news agencies' primary topics are well-represented in these graphs. This approach is adaptable to different datasets and can yield satisfactory results with new data. To the best of the author's knowledge, this method has not previously been utilized for the Persian language on Instagram. The newly collected data is openly available on GitHub for other researchers to explore. This method also enables the use of graph-based algorithms, including community detection.

**6. Future Works**

For future research, we propose exploring key phrase extraction and the identification of important sentences within texts. Additionally, expanding data sources to include diverse Persian news agencies can yield more comprehensive insights into evolving trends and topics in Iran. By implementing these suggestions, future studies can further enhance keyword extraction and network analysis, contributing to a deeper understanding of language semantics and trends on social media platforms like Instagram.

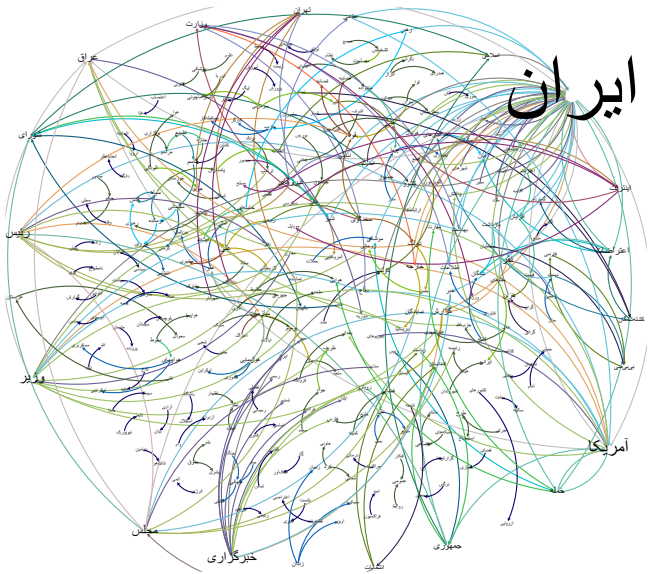

Figure. 11. Visualization of BBC Farsi Instagram Graph

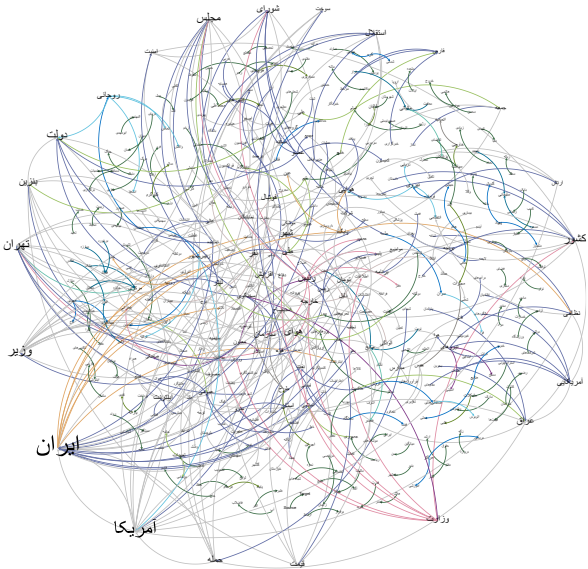

Figure. 12. Visualization of Instagram Graph of Fars News Agency

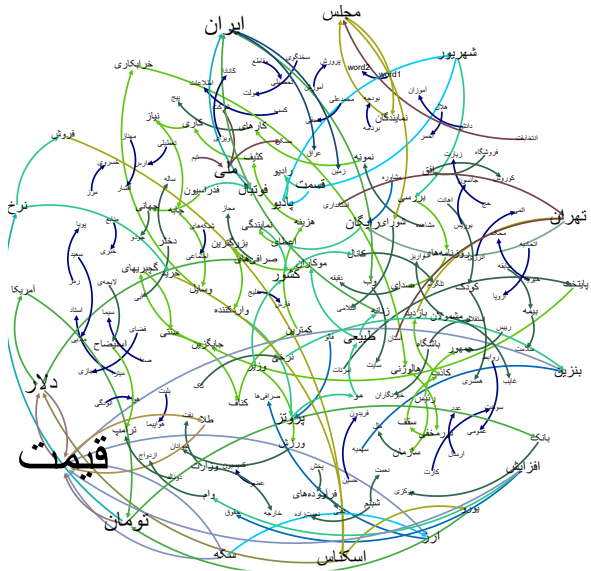

Figure. 13. KhabarFouri Instagram graph illustration

## Declarations

### Funding
This research did not receive any grant from funding agencies in the public, commercial, or non-profit sectors.

### Authors' contributions
MH: Study design, data modeling, statistical analysis, graph visualization, interpretation of the results and drafting the manuscript. MK: Data crawling and cleaning. BT: Drafting the manuscript, revision of the manuscript.

### Conflict of interest
The authors declare that no conflicts of interest exist.

## References

[1] A. Rejeb, K. Rejeb, A. Abdollahi, and H. Treiblmaier, "The big picture on Instagram research: Insights from a bibliometric analysis," *Telemat. Informatics*, vol. 73, no. December 2021, p. 101876, 2022, https://doi.org/10.1016/j.tele.2022.101876.

[2] C. Wartena, R. Brussee, and W. Slakhorst, "Keyword extraction using word co-occurrence," *Proc. - 21st Int. Work. Database Expert Syst. Appl. DEXA 2010*, IEEE, 2010, pp. 54–58, https://doi.org/10.1109/DEXA.2010.32.

[3] K. S. Hasan and V. Ng, "Automatic keyphrase extraction: A survey of the state of the art," *52nd Annu. Meet. Assoc. Comput. Linguist. ACL 2014 - Proc. Conf.*, vol. 1, pp. 1262–1273, 2014, https://doi.org/10.3115/v1/p14-1119.

[4] F. Liu, D. Pennell, F. Liu, and Y. Liu, "Unsupervised approaches for automatic keyword extraction using meeting transcripts," *NAACL HLT 2009 - Hum. Lang. Technol. 2009 Annu. Conf. North Am. Chapter Assoc. Comput. Linguist. Proc. Conf.*, IEEE, January 2009, pp. 620–628, https://doi.org/10.3115/1620754.1620845.

[5] S. Mirasdar and M. Bedekar, "Graph of Words Model for Natural Language Processing," in *Graph Learning and Network Science for Natural Language Processing*, CRC Press, 2023, pp. 1–20.

[6] M. Garg, *A survey on different dimensions for graphical keyword extraction techniques: Issues and Challenges*, vol. 54, no. 6, pp. 4731–4770, 2021. https://doi.org/10.1007/s10462-021-10010-6

[7] M. I. Fudolig, T. Alshaabi, M. V. Arnold, C. M. Danforth, and P. S. Dodds, "Sentiment and structure in word co-occurrence networks on Twitter," *Appl. Netw. Sci.*, vol. 7, no. 1, pp. 1–18, 2022, https://doi.org/10.1007/s41109-022-00446-2.

[8] R. Futrell, "Estimating word co-occurrence probabilities from pretrained static embeddings using a log-bilinear model," *C. 2022 - Work. Cogn. Model. Comput. Linguist. Proc. Work.*, pp. 54–60, 2022, https://doi.org/10.18653/v1/2022.cmcl-1.6.

[9] J. Gabín, M. E. Ares, and J. Parapar, "Keyword Embeddings for Query Suggestion," In *European Conference on Information Retrieval*, Cham: Springer Nature Switzerland, 2023, pp. 346–360, https://doi.org/10.1007/978-3-031-28244-7_22.

[10] D. A. Vega-Oliveros, P. S. Gomes, E. E. Milios, and L. Berton, "A multi-centrality index for graph-based keyword extraction," *Inf. Process. Manag.*, vol. 56, no. 6, p. 102063, 2019, https://doi.org/10.1016/j.ipm.2019.102063.

[11] S. K. Biswas, M. Bordoloi, and J. Shreya, "A graph based keyword extraction model using collective node weight," *Expert Syst. Appl.*, vol. 97, pp. 51–59, 2018, https://doi.org/10.1016/j.eswa.2017.12.025.

[12] W. Li, J. Xue, X. Zhang, H. Chen, and Z. Chen, "Word-Graph2vec: An efficient word embedding approach on word co-occurrence graph using random walk sampling," *arXiv preprint arXiv:2301.04312*, pp. 1–16. https://doi.org/10.48550/arXiv.2301.04312.

[13] H. Hettiarachchi, M. Adedoyin-olowe, J. Bhogal, and M. M. Gaber, "WhatsUp: An event resolution approach for co-occurring events in social media," *Inf. Sci. (Ny).*, vol. 625, pp. 553–577, 2023, https://doi.org/10.1016/j.ins.2023.01.001.

[14] R. Nazar and D. Lindemann, "Terminology extraction using co-occurrence patterns as predictors of semantic relevance," *In Proceedings of the Workshop on Terminology in the 21st century: many faces, many places*, 2022, pp. 26–29.

[15] F. Valentini, D. F. Slezak, and E. Altszyler, "The Undesirable Dependence on Frequency of Gender Bias Metrics Based on Word Embeddings," *arXiv preprint arXiv:2301.00792*, 2023.

https://doi.org/10.48550/arXiv.2301.00792.

[16] R. Mihalcea and P. Tarau, "TextRank: Bringing order into texts," *Proc. 2004 Conf. Empir. Methods Nat. Lang. Process. EMNLP 2004 - A Meet. SIGDAT, a Spec. Interes. Gr. ACL held conjunction with ACL 2004*, vol. 85, 2004, pp. 404–411.

[17] A. Xiong, D. Liu, H. Tian, Z. Liu, P. Yu, and M. Kadoch, "News keyword extraction algorithm based on semantic clustering and word graph model," *Tsinghua Sci. Technol.*, vol. 26, no. 6, pp. 886–893, 2021, https://doi.org/10.26599/TST.2020.9010051.

[18] J. Devlin, M.-W. Chang, K. Lee, and K. Toutanova, "Bert: Pre-training of deep bidirectional transformers for language understanding," *arXiv Prepr. arXiv1810.04805*, 2018. https://doi.org/10.48550/arXiv.1810.04805

[19] L. Chi and L. Hu, "ISKE: An unsupervised automatic keyphrase extraction approach using the iterated sentences based on graph method," *Knowledge-Based Syst.*, vol. 223, p. 107014, 2021, https://doi.org/10.1016/j.knosys.2021.107014.

[20] L. Gan, Z. Teng, Y. Zhang, L. Zhu, F. Wu, and Y. Yang, "SemGloVe: Semantic Co-Occurrences for GloVe From BERT," *IEEE/ACM Trans. Audio, Speech, Lang. Process.*, vol. 30, pp. 2696–2704, 2022, https://doi.org/10.1109/TASLP.2022.3197316.

[21] S. Anjali, N. M. Meera, and M. G. Thushara, "A Graph based Approach for Keyword Extraction from Documents," in *2019 Second International Conference on Advanced Computational and Communication Paradigms (ICACCP)*, 2019, pp. 1–4, https://doi.org/10.1109/ICACCP.2019.8882946.

[22] L. Zhao, Z. Miao, C. Wang, and W. Kong, "An Unsupervised Keyword Extraction Method based on Text Semantic Graph," in *2022 IEEE 6th Advanced Information Technology, Electronic and Automation Control Conference (IAEAC )*, 2022, pp. 1431–1436, https://doi.org/10.1109/IAEAC54830.2022.9929800.

[23] D. Brock, A. Khan, T. Doan, A. Lin, Y. Guo, and P. Tarau, "Textstar: a Fast and Lightweight Graph-Based Algorithm for Extractive Summarization and Keyphrase Extraction," *Proc. 20th Annu. Work. Australas. Lang. Technol. Assoc.*, no. 3, pp. 161–169, 2022, [Online]. Available: https://aclanthology.org/2022.alta-1.22.

[24] Y. Ying, T. Qingping, X. Qinzheng, Z. Ping, and L. Panpan, "A Graph-based Approach of Automatic Keyphrase Extraction," *Procedia Comput. Sci.*, vol. 107, no. Icict, pp. 248–255, 2017, https://doi.org/10.1016/j.procs.2017.03.087.

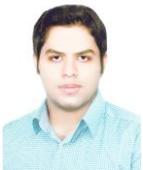

**Mohammad Heydari** received his B.Sc. degree in Computer Software Engineering, Technical and Vocational University, Tehan, Iran and received his M.Sc. degree in Information Technology Engineering, Tarbiat Modares University, Tehran, Iran. His research interests include Machine Learning, Natural Language Processing, and Knowledge Graphs.

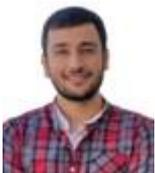

**Mohsen Khazeni** receivesd his B.Sc degree in Computer Software Engineering, Iran University of Science and Technology and received his M.Sc. degree in Information Technology Engineering, Tarbiat Modares University of Tehran. His research interests are Natural Language Processing, Deep Learning, and Social Network Analysis.

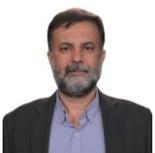

**Babak Teimourpour** received his B.Sc. degree in Industrial Engineering, Sharif University, Tehran, Iran and received his M.Sc. degree in SocioEconomic Systems Engineering, Institute for Research on Planning and Development, Tehran, Iran. Also, he received his Ph.D. in Industrial Engineering at Tarbiat Modares University, Tehran, Iran in 2010. His research interests include Data Mining, Social Network Analysis, and Bioinformatics.